\documentclass[
aps,%
12pt,%
final,%
prd,%
notitlepage,%
oneside,%
onecolumn,%
nobibnotes,%
nofootinbib,%
superscriptaddress,%
showpacs,%
centertags]%
{revtex4-1}

\pdfoutput=1

\usepackage[utf8]{inputenc} 
\usepackage[T2A,OT1]{fontenc} 
\usepackage[english]{babel} 
\usepackage{amssymb} 
\usepackage{mathtext} 
\usepackage{amsmath} 
\usepackage{mathtools} 
\usepackage{graphicx} 
\usepackage{epsfig}
\usepackage{hyperref}

\usepackage{tikz}
\usetikzlibrary{positioning,arrows}
\usetikzlibrary{decorations.pathmorphing}
\usetikzlibrary{decorations.markings}

\tikzset{
vector/.style={thick,double,draw=black, postaction={decorate},
    decoration={markings,mark=at position .6 with {\arrow[black,scale=0.8]{triangle 45}}}},
axial/.style={thick,double,densely dashed,draw=black, postaction={decorate},
    decoration={markings,mark=at position .6 with {\arrow[black,scale=0.8]{triangle 45}}}},
gluon/.style={decorate, draw=black,
    decoration={coil,aspect=0.3,segment length=5pt,amplitude=3pt}},
pseudo/.style={thick, dashed, draw=black, postaction={decorate},
    decoration={markings,mark=at position .6 with {\arrow[black]{triangle 45}}}},
scalar/.style={thick,draw=black, postaction={decorate},
    decoration={markings,mark=at position .6 with {\arrow[black]{triangle 45}}}},
cut/.style={very thick, densely dashed, draw=gray}
 }
\usepackage{upgreek} 
\input myunits.sty
\input mymath.sty
\newcommand{\figref}[1]{Fig.~\ref{#1}}

\newcommand{\appref}[1]{appendix~\ref{#1}}

\newcommand{\secref}[1]{section~\ref{#1}}

\newcommand{\eqnref}[1]{equation~(\ref{#1})}
\newcommand{\Eqnref}[1]{Equation~(\ref{#1})}
\begin{document}

\title{On the nature of $a_1(1420)$}
\author{M.~Mikhasenko}
\affiliation{Universit\"at Bonn, Helmholtz-Institut f\"ur Strahlen-
  und Kernphysik, 53115 Bonn, Germany}
\author{B.~Ketzer}
\affiliation{Universit\"at Bonn, Helmholtz-Institut f\"ur Strahlen-
  und Kernphysik, 53115 Bonn, Germany}
\author{A.~Sarantsev}
\affiliation{Universit\"at Bonn, Helmholtz-Institut f\"ur Strahlen-
  und Kernphysik, 53115 Bonn, Germany}
\affiliation{Petersburg Nuclear Physics Institute, Gatchina, Russia}

\begin{abstract}
The resonance-like signal with axial-vector quantum numbers
$J^{PC}=1^{++}$ at a mass of $1420\,\MeV$ and a width of $140\,\MeV$,
recently observed by the COMPASS and VES experiments in the
$f_0(980)\pi$ final state and tentatively called $a_1(1420)$, is
discussed. Instead of a genuine new meson, we interpret this signal as
a dynamical effect due to a singularity (branching point) in the
triangle diagram formed by the processes $a_1(1260) \to K^\star
\bar{K}$, $K^\star\to K\pi$, and $K \bar{K} \to f_0(980)$ (+ c.c).  
The  amplitude for this diagram is calculated. The
result exhibits a peak in the intensity with a sharp phase motion
with respect to the dominant $a_1(1260) \to \rho \pi$ $S$-wave
decay, in good agreement with the data. The branching ratio of
$a_1(1260)\to f_0(980)\pi$ via the triangle diagram is estimated and
compared to the dominant decay $a_1(1260)\to\rho\pi$. 
 
\end{abstract}
\maketitle

\section{Introduction}
\label{intro}

Understanding the quark interaction at low and intermediate energies
is one of the most challenging tasks of the theory of the strong
interactions. To create a theoretical non-perturbative approach or, at
least, to build a reliable model one should understand the nature of
strongly interacting particles and their excitation spectrum. The
classical quark model assumes that mesons are bound states of quarks
and antiquarks, and groups the low-mass states into nonets with the
same spin $J$, parity $P$ and charge-conjugation parity $C$, with
fixed mass differences between nonet members. The second assumption is
that the quark-antiquark interaction at large distances is governed by
a linearly rising potential which explains the phenomenon of quark
confinement and predicts the full spectrum of quark-antiquark excited
states. The success of the quark model is indisputable: most of the
known mesons correspond very well to the predicted scheme
\cite{Agashe:2014kda}.

However, it seems that the meson spectrum is notably richer than that
predicted by the quark model. There is a growing set of experimental
observations of resonance-like structures in partial waves with
quantum numbers which are forbidden for the quark-antiquark system or
situated at masses which can not be explained by the quark-antiquark
model, see e.g.\ \cite{Klempt:2007cp,Brambilla:2014jmp} and references
therein. 

Recently the COMPASS \cite{Adolph:2015ina,Paul:2013xra,Ketzer:2014raa} experiment
reported the observation of a small resonance-like signal with
axial-vector quantum numbers $I^G (J^{PC}) = 1^- (1^{++})$ in the
$f_0(980) \pi$ $P$-wave of the $\pi^-\pi^-\pi^+$ final state, produced
by diffractive scattering of a $190\,\GeV$ $\pi^-$ beam on a proton
target.  The signal was also confirmed by the VES experiment
\cite{Khokhlov:2014nha} in the $\pi^-\pi^0\pi^0$ final state. In both
experiments, the three-pion final states
were analyzed using a two-step partial-wave analysis (PWA)
technique. In the first step the data were grouped in small bins of
$3\pi$-invariant mass and momentum transfer.  The isobar model was
employed to parametrize possible decays to three final pions.  An
isolated, relatively narrow peak was found in the intensity of the
$1^-1^{++}\,f_0\pi\,P$-wave at a mass around $1.4\,\GeV$, accompanied
by
a sharp phase motion of this wave relative to other known resonances,
with a phase variation exceeding $180^\circ$.
In the second step of the COMPASS and VES analyses the spin-density
matrix resulting from the first step was fitted with a model including
Breit-Wigner resonances and background contributions.  The new signal
was described rather well with a hitherto unknown resonance,
which was tentatively called $a_1(1420)$ with a mass $M_{a_1} \approx 1.42\,$GeV
and width $\Gamma_{a_1}\approx 0.14\,$GeV. The interpretation of this
signal as a new state in the framework of the quark model is
difficult. It cannot be a radial excitation of $a_1(1260)$ which is
expected to have a mass above $1650\,$MeV. It is also not expected
that the radial excitation has a width which is much smaller than the
one of the ground state.  Therefore, this signal is to be considered
either as a strong candidate for a four-quark bound state or a
meson-meson molecular bound state or to be explained as a some
dynamical effect resulting from multi-particle interaction.

In the present paper we show that a signal of comparable strength,
including the rapid phase motion, can be expected by the opening of
the $K^{\star}\bar{K}$ decay channel for the isovector $a_1(1260)$,
and the re-scattering of the kaons. 
There are two isospin combinations of
intermediate particles:
\begin{itemize}
\item[(i)] $a_1^-(1260) \to K^{\star 0} K^- \to \pi^- K^+ K^- \to \pi^-
  f_0$,
\item[(ii)] $a_1^-(1260) \to K^{\star -} K^0 \to \pi^- \bar{K}^0 K^0 \to
  \pi^- f_0$.
\end{itemize}
The corresponding
triangle diagram has a logarithmic singularity on the tail of the wide
$a_1(1260)$ which is due to a very
peculiar kinematic situation,  
in
which all intermediate particles are almost on their mass shell,
causing a resonance-like effect. 

Recently, Wu et al.\ \cite{Wu:2011yx,Wu:2012pg} showed that 
the same triangle singularity in the isospin-0 channel can account for the
anomalously large isospin violation effects observed by BESIII for
$\eta(1407/1475)$ and $f_1(1420)\to f_0(980)\pi \to 3\pi$ 
\cite{BESIII:2012aa}.  
A similar dynamic mechanism was suggested earlier by Achasow and Kozhevnikov
\cite{Achasov:1989ma} to explain the resonance-like signal observed in
the $\phi \pi^0$ mass spectrum of the reaction $\pi^- p\to \phi\pi^0
n$ \cite{Bityukov:1986yd}
by the decay of $\rho(1700)\to K^\star\bar{K}$ and rescattering of
$K\bar{K}\to\phi$.  
Triangle singularities are currently also being discussed in the
context of the newly discovered XYZ quarkonium peaks 
\cite{Szczepaniak:2015eza}.  





Our paper is organized as follows. 
In sections~\ref{sec:kinematics} to \ref{sec:triangle}, we only discuss
the triangle diagram for process (i) with intermediate particles
$(K^{\star 0},K^+,K^-)$, the calculation  
for process (ii) proceeds analogously. 
In \secref{sec:kinematics} the kinematic conditions for the appearance
of the triangle singularity are analyzed.  
The amplitude for the triangle process is calculated in the following
two sections.  In \secref{sec:imaginary} we first present an approach
to calculate the imaginary part of the amplitude making use of
Cutkosky cutting rules and the calculation of discontinuities. This
method helps to understand the structure of the amplitude
singularities.  In \secref{sec:triangle} we then use 
an effective
Lagrangian approach \cite{Meissner:1987ge} to
calculate the full amplitude, i.e. the real and imaginary parts,
needed to predict the phase motion. In both sections, we start with
the case of scalar particles to illustrate the underlying physics. 
In scalar theory the behavior of the amplitude is 
$\propto\log(s-E_1^2)$ near the singularity.
Since the amplitude behavior could be different in scalar theory and
in interactions of particles with spin \cite{Achasov:1989ma}, then the
realistic 
situation for particles with spin and finite width is calculated. 
We show that the singularity is removed only by including the finite
width of unstable particles and gives
a contribution $\sim \log \Gamma_{K^\star}$. 


In \secref{sec:reaction} we estimate the branching ratio of
$a_1(1260) \to f_0(980)\pi^-$, now including both isospin combinations, 
relative to its dominant decay channel 
$\rho\pi^-$, and compare the signal we expect for the triangle diagram
to experimental values as reported in \cite{Ketzer:2014raa}.


\section{Kinematic conditions for triangle singularity}
\label{sec:kinematics}
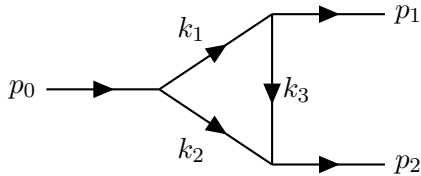
\begin{figure}[ht]
  \centering
  \begin{tikzpicture}[node distance=1cm and 1.5cm]
    \coordinate[label=left:{$p_0$}] (a1); \coordinate[right=of a1]
    (a2); \coordinate[above right=of a2] (b1); \coordinate[below
    right=of a2] (b2); \coordinate[right=of b1,label=right:$p_1$]
    (c1); \coordinate[right=of b2,label=right:$p_2$] (c2);

    \draw[scalar] (a1) -- (a2); \draw[scalar] (a2) -- node[label=above
    left:$k_1$ ] {} (b1); \draw[scalar] (a2) -- node[label=below
    left:$k_2$ ] {} (b2); \draw[scalar] (b1) -- node[label=right:$k_3$
    ] {} (b2); \draw[scalar] (b1) -- (c1); \draw[scalar] (b2) -- (c2);
  \end{tikzpicture}
  \caption{The process $0 \to 1+2$ for particles with 4-momenta $p_0$,
    $p_1$, $p_2$, proceeding via a triangle diagram 
    with intermediate particle momenta $k_1$, $k_2$, $k_3$.}
  \label{fig:triangle}
\end{figure}
\begin{figure}[ht]
  \centering
  \includegraphics[width=0.4\textwidth]{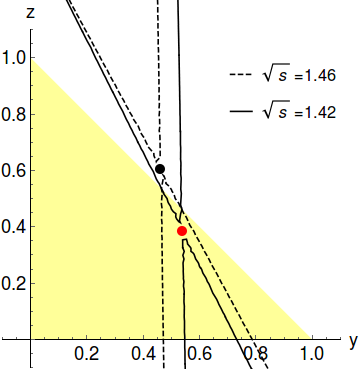}
  \caption{Diagram illustrating the positions of singularities of the
    triangle diagram shown in \figref{fig:triangle} for the Feynman
    variables $y,\,z$. The light-gray triangle is the
    kinematically allowed region. 
    The black dots are pinch singularities,
    corresponding to $\sqrt{s}=E_1$ (lower dot) and $\sqrt{s}=E_2$
    (upper dot), respectively. The curves are solutions of
    $\Delta_{yz}+m_1^2(1-y-z) 
    =0$ (see equations~\ref{eq:scalar_case} and \ref{eq:Delta}) for
    $\sqrt{s}=E_1$ (solid line) and $\sqrt{s}=E_2$ (dashed 
    line).}
  \label{fig:feynman_var}
\end{figure}

It is well known that logarithmic singularities arise in processes
which proceed via the triangle diagram shown in \figref{fig:triangle}. 
As it was shown 
by a general analysis of 
singularities in scalar theory \cite{Landau:1959fi}, the amplitude
behavior near the branching point 
of a cut is $\propto\log(s-s_0)$, where $s$ is an external invariant. The position of the singularity $s_0$ can be obtained from the  
simple condition that all intermediate particles are on mass shell and
collinear to each other. It is  
given by the system of 
Lorentz-invariant Landau equations:
\begin{equation} \label{eq:sing_cond} \left\{
    \begin{array}{lll}
      k_i^2 = m_i^2, && i=1\dots 3, \\
      x\, k_{1\mu} - y\, k_{2\mu} + z\, k_{3\mu} = 0, && x,y,z \in [0,1], \\
      x+y+z=1,
    \end{array}
  \right.
\end{equation}
with $k_i$ and $m_i$ the 4-momenta and masses of intermediate
particles, respectively, and $x,y,z$ the so-called Feynman
parameters.     
The system of equations~(\ref{eq:sing_cond}) for $x,y,z$ is
overdetermined, so it is solvable only in  
exceptional cases. 
For the special case of the decay of $a_1^-(1260)$ to $f_0(980) \pi^-$
through intermediate 
particles 
$(K^{\star-}, \bar{K}^0, K^0)$ or
$(K^{\star 0}, K^+, K^-)$, and neglecting the finite width of the
$f_0(980)$, the external momenta 
$p_1$ and $p_2$ depend only on $s=p_0^2$ (see \figref{fig:triangle}
for the definition of $p_i$).
Using kinematical relations between internal and external momenta
it can be shown that the system~(\ref{eq:sing_cond}) has solutions
only if 
$\sqrt{s} = E_{1,2}$, where $E_1 = 1.42\,$GeV, $E_2 = 
1.46\,$GeV. These pinch singularities are shown as dots in
\figref{fig:feynman_var}. As can be seen, the conditions
$x,y,z\in[0,1]$ are satisfied only for the first solution. 

Here we give a simple kinematic explanation for the appearance of the
singularity. The initial state 
$a_1(1260)$ with $J^{PC}=1^{++}$ can decay to real $K^\star \bar{K}+c.c.$
starting from the threshold 
$\sqrt{s} = 1.39\,$GeV. Then the $K^\star$ decays to real $K$ and $\pi$.
Note that the $K$ from $K^\star$ decay can go to the same direction as
the $\bar{K}$, the ratio of velocities of $\bar{K}$ and $K$ is a 
function of $\sqrt{s}$ as well as of the invariant mass of $K$ and
$\bar{K}$. The invariant mass of $\bar{K}$ $K$ going in 
the same direction is equal to the mass of $f_0$ only if $\sqrt{s} =
E_{1,2}$, but for $E_2$ the $\bar{K}$ is faster than $K$ and thus the
$K$ cannot 
catch up the $\bar{K}$ to form $f_0$. Only for the solution $E_1$ do
the $K$ and $\bar{K}$ proceed in the 
same direction with the same velocity with an invariant mass equal to
that of the $f_0$. 

The kinematics discussed here demonstrates a very peculiar situation
in the decay of the $a_1(1260)$ to $K^\star \bar{K}+c.c$: just above
the two-body threshold, the re-scattering  
in the triangle can happen with particles almost on mass shell. 


\clearpage

\section{Imaginary part of the amplitude}
\label{sec:imaginary}

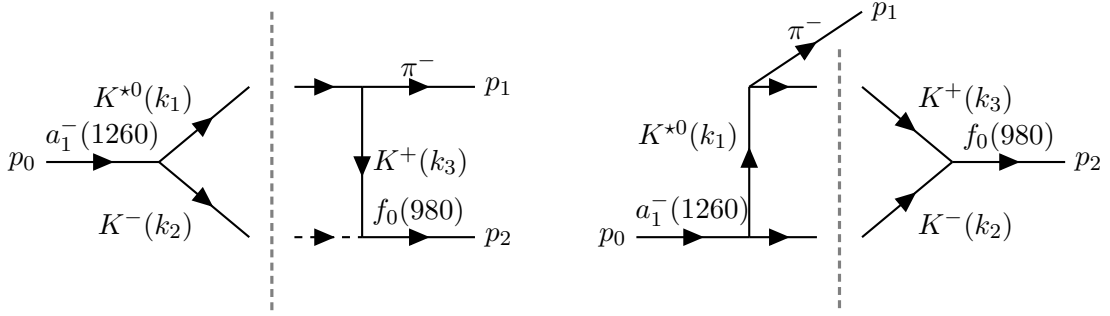
\begin{figure}[ht]
  \centering
  \begin{tikzpicture}[node distance=1cm and 1.5cm]
    \coordinate[label=left:{$p_0$}] (a1);
    \coordinate[right=of a1] (a2); \coordinate[above right=of
    a2,xshift=-3mm] (b1); \coordinate[below right=of a2,xshift=-3mm]
    (b2); \coordinate[above right=of a2,yshift=+1cm] (cb1);
    \coordinate[below right=of a2,yshift=-1cm] (cb2);
    \coordinate[above right=of a2,xshift=+3mm] (bb1);
    \coordinate[below right=of a2,xshift=+3mm] (bb2);
    \coordinate[right=of b1] (d1); \coordinate[right=of b2] (d2);
    \coordinate[right=of d1,label=right:$p_1$] (c1);
    \coordinate[right=of d2,label=right:$p_2$] (c2);

    \draw[scalar] (a1) --
    node[label=above:\textcolor{black}{$a_1^-(1260)$}] {} (a2);
    \draw[scalar] (a2) -- node[label=above
    left:\textcolor{black}{$K^{\star 0}(k_1)$} ] {} (b1); \draw[scalar]
    (a2) -- node[label=below left:\textcolor{black}{$K^-(k_2)$} ] {}
    (b2); \draw[scalar] (d1) --
    node[label=right:\textcolor{black}{$K^+(k_3)$} ] {} (d2);
    \draw[scalar] (bb1) -- (d1); \draw[pseudo] (bb2) -- (d2);
    \draw[scalar] (d1) -- node[label=above:\textcolor{black}{$\pi^-$}
    ] {} (c1); \draw[scalar] (d2) --
    node[label=above:\textcolor{black}{$f_0(980)$} ] {} (c2);
    \draw[cut] (cb1) -- (cb2);
  \end{tikzpicture}
  \qquad
  \begin{tikzpicture}[node distance=1cm and 1.5cm]
    \coordinate[label=right:$p_2$] (a1); \coordinate[left=of a1] (a2);
    \coordinate[above left=of a2,xshift=+3mm] (b1); \coordinate[below
    left=of a2,xshift=+3mm] (b2); \coordinate[above left=of
    a2,yshift=+5mm] (cb1); \coordinate[below left=of a2,yshift=-1cm]
    (cb2); \coordinate[above left=of a2,xshift=-3mm] (bb1);
    \coordinate[below left=of a2,xshift=-3mm] (bb2);
    \coordinate[left=of b1] (d1); \coordinate[left=of b2] (d2);
    \coordinate[above right=of d1,label=right:$p_1$] (c1);
    \coordinate[left=of d2,label=left:{$p_0$}] (c2);

    \draw[scalar] (a2) --
    node[label=above:\textcolor{black}{$f_0(980)$}] {} (a1);
    \draw[scalar] (b1) -- node[label=above
    right:\textcolor{black}{$K^+(k_3)$} ] {} (a2); \draw[scalar] (b2)
    -- node[label=below right:\textcolor{black}{$K^-(k_2)$} ] {} (a2);
    \draw[scalar] (d2) --
    node[label=above left:\textcolor{black}{$K^{\star 0}(k_1)$} ] {} (d1);
    \draw[scalar] (d1) -- (bb1); \draw[scalar] (d2) -- (bb2);
    \draw[scalar] (d1) -- node[label=above:\textcolor{black}{$\pi^-$}
    ] {} (c1); \draw[scalar] (c2) --
    node[label=above:\textcolor{black}{$a_1^-(1260)$} ] {} (d2);
    \draw[cut] (cb1) -- (cb2);
  \end{tikzpicture}
  \caption{Two possible cuts which contribute to the imaginary part of
   the matrix element of the process $a_1(1260) \to f_0(980) \pi^-$.}
  \label{fig:cuts.scalar}
\end{figure}

In order to understand the structure of the amplitude, we first
consider the imaginary part only, 
based on discontinuities. The technique was developed by Cutkosky \cite{Cutkosky:1960sp}, is described e.g. in the Gribov lectures
\cite{Gribov:2009zz}, and was successfully applied by Achasov and
Kozhevnikov for similar process, $\rho' \to \phi \pi$
\cite{Achasov:1989ma}.  The imaginary part of the amplitude $\mathbb{M}$ of the diagram in \figref{fig:triangle} is related to the discontinuity across the cuts shown
in \figref{fig:cuts.scalar} by
\begin{equation} \label{eq:ImDisc} 
\text{Im}\, \mathbb{M}_{a_1\to f_0 \pi} = \frac{1}{2} \left( \text{Disc}_{K^\star K} + \text{Disc}_{K\bar{K}}
  \right)\enspace.
\end{equation}
To calculate the discontinuities, we use the following expression
\begin{equation}
  \text{Disc} = \int \prod_{\text{cut}} \frac{\mathrm{d}^3 k_i}{(2\pi)^3 2E_i^k} \times 
  \left( \sum_{\text{polarization}}\mathbb{M}_1 \cdot \mathbb{M}_2^\star \right)
  \times (2\pi)^4 \delta^4(\text{mom. cons.})\enspace,
\end{equation}
where $\mathbb{M}_{1,2}$ are matrix elements for processes on the left and right hand side of the cutting line, respectively (see \figref{fig:cuts.scalar}). We are calling particles which are crossed by cut line as cut particles.
The integration is over all momentum space for cut particles, i.e. $k_i$ are momenta of cut particles, $E_i^k$ are the corresponding energies. If a cut particle has spin we sum over all possible polarizations.

\subsection{Simple model with scalar intermediate particles}
\label{sec:imaginary.scalar}

For the case of scalar intermediate particles, the expressions for the discontinuities are:
\begin{equation} \label{eq:Disc12_sc_def} 
\text{Disc}_{K^\star K}^{(\text{sc})} = g^3\int
 \frac{\mathrm{d}^3 k_1}{(2\pi)^3 2E_1^k} \frac{\mathrm{d}^3 k_2}{(2\pi)^3 2E_2^k}
  \times \frac{1}{m_3^2-k_3^2 +
    i\epsilon} \times (2\pi)^4 \delta^4(p_0-k_1-k_2)\enspace,
\end{equation}
\begin{equation} \label{eq:Disc23_sc_def} 
\text{Disc}_{K\bar{K}}^{(\text{sc})} = g^3\int
 \frac{\mathrm{d}^3 k_2}{(2\pi)^3 2E_2^k} \frac{\mathrm{d}^3 k_3}{(2\pi)^3 2E_3^k}
  \times \frac{1}{m_1^2-k_1^2 - i\epsilon} \times (2\pi)^4 \delta^4(k_3+k_2-p_2)\enspace.
\end{equation}

Here, the products of matrix elements $\mathbb{M}_1 \cdot \mathbb{M}_2^\star$ are given by the coupling constants at the three vertices, which are set to $g$, and the propagator, which is a function of the angle between $\vec k_1$ and $\vec p_1$ in \eqnref{eq:Disc12_sc_def} and a function of the angle between $\vec k_2$ and $\vec p_1$ in \eqnref{eq:Disc23_sc_def}. For both discontinuities, the cut particles ($K^\star$, $K^-$ and $K^+$, $K^-$, respectively) are set on their mass shells.  
Integration with delta function in \eqnref{eq:Disc23_sc_def} is performed
in the $f_0$ rest frame.  After carrying out the integration we arrive at the following
expression:
\begin{equation} \label{eq:Disc_sc}
\text{Im}\, \mathbb{M}_{a_1\to f_0 \pi}^{(\text{sc})} = \frac{g^3}{16\pi} \left[
  \frac{1}{2|\vec{p}\,| \sqrt{s} } \log \frac{\tilde{A}+1 + i\epsilon}{\tilde{A}-1+i\epsilon} + 
  \frac{1}{2|\vec{p}\,'| M_1} \log \frac{\tilde{C}+1-i\epsilon}{\tilde{C}-1-i\epsilon}
  \right]\enspace,
\end{equation}
where the coefficients $\tilde{A}$, $\tilde{C}$ originate from the propagators,
\begin{equation}
  \tilde{A} = (m_3^2-m_1^2-M_1^2+2E_1^p E_1^k)/(2|\vec k| |\vec p\,|)\enspace,
\end{equation}
\begin{equation}
  \tilde{C} = (m_1^2 - s - m_2^2+2E_0' E_2^{k\prime})/(2|\vec{k}\,'| |\vec{p}\,'|)\enspace.
\end{equation}
Here, $M_i^2 = p_i^2$, 
$| \vec{k} | = \lambda^{1/2}(s,m_1^2,m_2^2)/(2\sqrt{s})$, $| \vec{p}\, | =
\lambda^{1/2}(s,M_1^2,M_2^2)/(2\sqrt{s})$ are the momenta of the corresponding particles in $a_1$ rest frame, with the K\"allen function 
\begin{equation}
\lambda(x,y,z) = x^2+y^2+z^2-2(xy+yz+zx)\enspace.
\end{equation}
$E_1^k = (s+m_1^2-m_2^2)/(2\sqrt{s})$, $E_1^p = (s+M_1^2-M_2^2)/(2\sqrt{s})$.
The values with prime are calculated at the $f_0$ rest frame: $|\vec{k}'| = \lambda^{1/2}(M_2^2,m_2^2,m_3^2)/(2M_2)$, $|\vec{p}\,'| =
\lambda^{1/2}(s,M_1^2,M_2^2)/(2M_2)$ are the momenta of $K^+$ ($K^-$) and
$\pi^-$ ($a_1^-$). The corresponding expressions for energy are
$E_2^{k\prime} = (M_2^2+m_2^2-m_3^2)/(2M_2)$, $E_0' =
(s+M_2^2-M_1^2)/(2M_2)$.
The imaginary parts of the expressions (\ref{eq:Disc12_sc_def}), (\ref{eq:Disc23_sc_def}) compensate each other and the \eqnref{eq:Disc_sc} is real.

The imaginary part of the amplitude $\text{Im} \,\mathbb{M}_{a_1 \to f_0 \pi}^{(\text{sc})}(s)$ as well as the contributions from the individual discontinuities are shown in \figref{fig:im} by dashed lines.
One can clearly see singularity $\sqrt{s} = E_1$ and $E_2$ values. 
The singularity in $E_2$ is out of kinematically allowed
region of reaction, so the sum of the two discontinuities is smooth at $E_2$.
One can also notice that the imaginary part is not zero below the threshold
of $K^\star K$ threshold. Here, the contribution comes from Disc$_{K\bar{K}}$, 
because the mass $m_{f_0}$ is above $2m_{K^\pm}$ threshold. Of course, taking into account the real shape of $K^\star$ and $f_0$ will make the amplitude smoother, as shown in \secref{sec:triangle.kstar}, but the effect of the singularity at $\sqrt{s}=E_1$ will remain. This conclusion will also not change when the spin of the particles is taken into account, as will be shown in the next section.

\subsection{Realistic case: VPP intermediate particles}
\label{sec:imaginary.vpp}

In reality, the particles involved in the process carry quantum numbers different from the scalar particles used in the previous section. 
The $a_1(1260)$ with axial-vector quantum numbers $J^P=1^+$ decays to
$K^\star \bar{K}$ with vector and pseudoscalar quantum numbers,
respectively. The $K^\star$ decays to two pseudoscalars, $K\pi$.
The Feynman rules for the hadronic vertices which we use are given in \appref{sec:adx.feynmann_rules}.

The expressions for the discontinuities are:
\begin{equation} \label{eq:Disc12_def} 
\text{Disc}_{K^\star \bar{K}}^{(\text{vpp})} = g^3\int
  \frac{\mathrm{d}^3 k_1}{(2\pi)^3 2E_1^k} \frac{\mathrm{d}^3 k_2}{(2\pi)^3 2E_2^k}
  \times \frac{ \varepsilon_{0\mu} \left( g^{\mu\nu} - \frac{k_1^\mu
        k_1^\nu}{m_1^2} \right) (p_1-k_3)_\nu }{m_3^2-k_3^2 +
    i\epsilon} \times (2\pi)^4 \delta^4(p_0-k_1-k_2),
\end{equation}
\begin{equation} \label{eq:Disc23_def} 
\text{Disc}_{K\bar{K}}^{(\text{vpp})} = g^3\int
  \frac{\mathrm{d}^3 k_2}{(2\pi)^3 2E_2^k } \frac{\mathrm{d}^3 k_3}{(2\pi)^3 2E_3^k }
  \times \frac{ \varepsilon_{0\mu} \left( g^{\mu\nu} - \frac{k_1^\mu
        k_1^\nu}{m_1^2} \right) (p_1 - k_3)_\nu }{m_1^2-k_1^2 -
    i\epsilon} \times (2\pi)^4 \delta^4(k_3+k_2-p_2).
\end{equation}
Here $\varepsilon_0$ is a polarization vector of the $a_1$ state, we use notation shown in \figref{fig:cuts.scalar}.

\begin{figure}[h]
  \centering
  \includegraphics[width=0.7\textwidth]{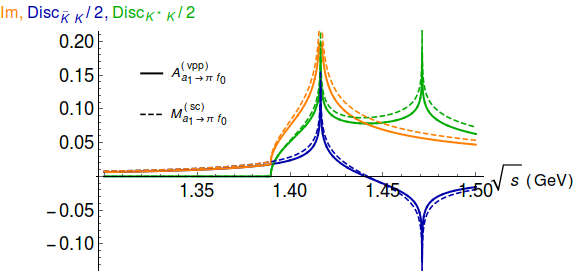}
  \caption{Energy dependence of $\text{Im} \,\mathbb{M}_{a_1 \to f_0 \pi}^{(\text{sc})}(s)$ and $\text{Im} \,\mathbb{A}_{a_1 \to f_0 \pi}^{(\text{vpp})}(s)$ (dashed and full lines, respectively). 
The contributions of the discontinuities $K^\star \bar{K}$ and $K\bar{K}$ are shown by green and blue lines, respectively. }
  \label{fig:im}
\end{figure}

After integration, we have:
\begin{multline} \label{eq:Disc12}
  \text{Disc}_{K^\star \bar{K}}^{(\text{vpp})} = g^3\frac{1}{8\pi} \frac{2|\vec{k}|}{\sqrt{s}} \, 
  (\varepsilon_{0} (p_1-p_2)) \times \\
  \times \frac{M_1^2+m_1^2-m_3^2}{4 m_1^2\vec p_1^2} \left[ 1 + \left(
      -\frac{|\vec p |}{|\vec k|}\frac{m_1^2}{M_1^2+m_1^2-m_3^2} +
      \frac{\tilde{A}}{2} \right) \log \frac{\tilde{A}-1+
      i\epsilon}{\tilde{A}+1+ i\epsilon} \right]\enspace,
\end{multline}
\begin{multline} \label{eq:Disc23}
  \text{Disc}_{K\bar{K}}^{(\text{vpp})} = -g^3\frac{1}{8\pi} \frac{2|\vec k'|}{M_2} \, 
  (\varepsilon_{0} (p_1-p_2)) \times  \\
 \times \frac{M_1^2+m_1^2-m_3^2}{4 m_1^2 \vec p\,'^2} \frac{E_0'}{M_2}
  \left[ 1+\left( \frac{|\vec p\,'|}{|\vec k'|}\frac{2 m_1^2 M_2 -
        (M_1^2+m_1^2-m_3^2) E_2^{k\prime}}{(M_1^2+m_1^2-m_3^2)E_0'} +
      \frac{\tilde{C}}{2} \right)
    \log\frac{\tilde{C}-1-i\epsilon}{\tilde{C}+1-i\epsilon}
  \right]\enspace.
\end{multline}
The notations could be found in the previous section.
The $P$-wave from $K^\star$ decay is propagated to the $f_0 \pi$
$P$-wave. So we have a factor $(\varepsilon_{0} (p_1-p_2))$ in the final
expression for the imaginary part of the matrix element. We separate it to
compare the result with the scalar case: 
\begin{equation} 
  \mathbb{M}_{a_1\to f_0 \pi}^{(\text{vpp})} = g^3 \, \mathbb{A}_{a_1\to
    f_0 \pi}^{(\text{vpp})} (\varepsilon_{0} (p_1-p_2))\enspace.
  \label{eq:relationMA}
\end{equation}
$\mathbb{A}_{a_1\to f_0 \pi}^{(\text{vpp})}$ is plotted in \figref{fig:im} together with the result from the scalar theory. The two results are very similar. 

\clearpage

\section{Full amplitude for $a_1^-(1260)\to f_0(980) \, \pi^-$ via
  $K^{\star 0}K^+K^-$ triangle}
\label{sec:triangle}



\begin{figure}[htb]
  \centering
  \begin{tikzpicture}[node distance=1cm and 1.5cm]
    \coordinate[label=left:{$p_0$}] (a1);
    \coordinate[right=of a1] (a2); \coordinate[above right=of a2]
    (b1); \coordinate[below right=of a2] (b2); \coordinate[right=of
    b1,label=right:$p_1$] (c1); \coordinate[right=of
    b2,label=right:$p_2$] (c2);

    \draw[scalar] (a1) -- node[label=above:$a_1^-(1260)$] {} (a2);
    \draw[scalar] (a2) -- node[label=above left:$K^{\star 0}(k_1)$ ] {}
    (b1); \draw[scalar] (a2) -- node[label=below left:$K^-(k_2)$ ] {}
    (b2); \draw[scalar] (b1) -- node[label=right:$K^+(k_3)$ ] {} (b2);
    \draw[scalar] (b1) -- node[label=above:$\pi^-$ ] {} (c1);
    \draw[scalar] (b2) -- node[label=below:$f_0(980)$ ] {} (c2);
  \end{tikzpicture}
  \caption{$a_1(1320) \to f_0(980) \pi^-$ triangle diagram.}
  \label{fig:diagram.scalar}
\end{figure}
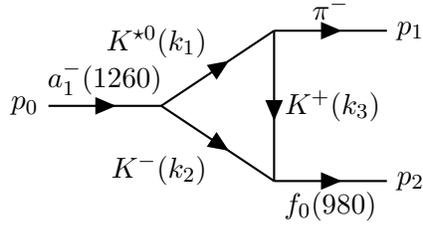

After the calculation of the imaginary part of the amplitude based on
discontinuities we proceed now to the calculation of the full
amplitude for the triangle diagram shown in \figref{fig:diagram.scalar} using
Feynman rules for hadronic processes in an effective Lagrangian
approach \cite{Meissner:1987ge} (see \appref{sec:adx.feynmann_rules}
for the parameterization of vertices).  
As in the previous section, we start from the simple case of scalar particles, and generalize to particles with spin in
\secref{sec:triangle.vpp}. 

\subsection{Simple model with scalar intermediate particles}
\label{sec:triangle.scalar}
\begin{figure}[tbp]
  \centering
  \includegraphics[width=0.49\textwidth]{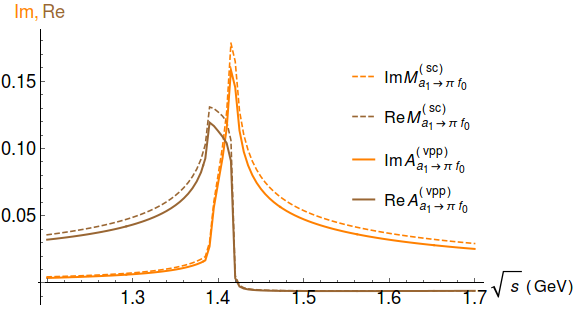} \
  \includegraphics[width=0.49\textwidth]{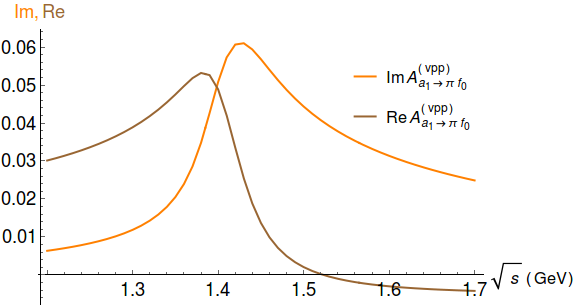} 
  \caption{(Left) Real (brown) and imaginary (orange) parts of
    $\mathbb{A}_{a_1\to\pi f_0}^{(\text{vpp})}(s)$ and
    $\mathbb{M}_{a_1\to\pi f_0}^{(\text{sc})}(s)$, respectively.   
    (Right) Real (brown) and imaginary (orange) parts of
    $\mathbb{A}_{a_1\to\pi f_0}^{(\text{vpp})}(s)$ when the
    finite width of $K^\star$ is taken into account in VPP case.} 
  \label{fig:ReIm}
\end{figure}

In case of vertices involving scalar particles only the matrix element
for the triangle diagram in \figref{fig:diagram.scalar} is  
\begin{equation}
\mathbb{M}_{a_1\to f_0 \pi}^{(\text{sc})} = g^3 \int\frac{\mathrm{d}^4 k_1}{(2\pi)^4 i}\, \frac{1}{(m_1^2 - k_1^2 - i\epsilon)(m_2^2 - (p_0-k_1)^2 - i\epsilon)(m_3^2 - (k_1-p_1)^2 - i\epsilon)}\enspace,
\label{eq:Msc}
\end{equation}
We calculate the integral using the standard technique of Feynman parameters and Wick rotation, 
which is described in more detail in \appref{sec:adx.integral}:
\begin{equation} \label{eq:scalar_case}
\mathbb{M}_{a_1\to\pi f_0}^{(\text{sc})} =
  \frac{g^3}{16\pi^2} \int_0^1 \mathrm{d}y \int_0^{1-y} \mathrm{d}z
   \frac{1}{\Delta_{yz}+m_1^2(1-y-z)-i\epsilon}\enspace,
\end{equation}
where $\Delta_{yz}$ is given by 
\begin{equation} \label{eq:Delta}
\Delta_{yz} = ym_2^2 + zm_3^2 - y(1-y-z)p_0^2 - z(1-z-y)p_1^2 - yz p_2^2\enspace.
\end{equation}
\Eqnref{eq:scalar_case} is evaluated numerically and the real and 
imaginary parts are shown in \figref{fig:ReIm} (left panel). 
If the widths of all intermediate particles are set to zero, the 
detailed structure of the amplitude becomes apparent.   
The imaginary part starts to grow rapidly from threshold
$\sqrt{s_{\text{th}}} = m_{K}+m_{K^\star}$ and goes to infinity when
$\sqrt{s} = E_1$. It exactly reproduces our result from
\secref{sec:imaginary} shown in \figref{fig:im}.   
The real part has a cusp at the threshold, 
then sharply drops below zero at $\sqrt{s} = E_1$ and becomes stable
for higher values of $\sqrt{s}$.

\subsection{Realistic case: VPP intermediate particles}
\label{sec:triangle.vpp}



For the realistic case of vector and pseudoscalar intermediate
particles the expression for the matrix element is
\begin{equation} \label{eq:Mvpp}
\mathbb{M}_{a_1\to\pi f_0}^{(\text{vpp})} = 
 g^3 \int \frac{\mathrm{d} k_1^4 }{(2\pi)^4\, i} \frac{\varepsilon_{0\mu} \left( g^{\mu\nu} - \frac{k_1^\mu k_1^\nu}{k_1^2} \right) (p_{1} - k_3)_\nu}
 {(m_1^2 - k_1^2 - i\epsilon)(m_2^2 - (p_0-k_1)^2 - i\epsilon)(m_3^2 - (k_1-p_1)^2 - i\epsilon)}\enspace.
\end{equation}

We can apply the same
procedure as in \secref{sec:triangle.scalar}, introducing 
Feynman parameters and performing a Wick rotation, and then calculate
the resulting integral numerically. 
The details of the calculation are shown in
\appref{sec:adx.integral.1}. Using the relation between
$\mathbb{A}_{a_1\to\pi f_0}^{(\text{vpp})}$ and $\mathbb{M}_{a_1\to\pi
  f_0}^{(\text{vpp})}$ given in \eqnref{eq:relationMA}, the 
result is 
\begin{multline} \label{eq:vvp_case} 
\mathbb{A}_{a_1\to\pi f_0}^{(\text{vpp})} = \frac{1}{16\pi^2} \int_0^1 \mathrm{d}y \int_0^{1-y} \mathrm{d} z  \frac{1}{\Delta_{yz} + m_1^2(1-y-z)-i\epsilon} + \\
+ \frac{1}{16\pi^2} \int_0^1 \mathrm{d}y \int_0^{1-y} \mathrm{d}z 
\int_0^{1-y-z} \mathrm{d}x \left( \frac{yz(p_0\cdot p_1) + z^2 p_1^2}{(\Delta_{yz} + m_1^2 x-i\epsilon)^2} - \frac{1/4}{ \Delta_{yz} + m_1^2 x-i\epsilon }\right)\enspace.
\end{multline}
The real and imaginary parts of $\mathbb{A}_{a_1\to\pi
  f_0}^{(\text{vpp})}$ are compared to the scalar case in
\figref{fig:ReIm} (left). For both real and imaginary parts, the result for
the VPP case is very similar to the scalar one, as was already shown for the imaginary part in 
\secref{sec:imaginary}.  

\subsection{Corrections to $K^\star\to K\pi$ vertex}
\label{sec:triangle.kstar}
There are two additional corrections to be taken into account in order to
arrive at a realistic estimate of the triangle amplitude: 
\begin{enumerate}
\item Finite widths of intermediate particles. Until now we have
  assumed that the 
  particles in the loop are stable ($\epsilon \to 0$). While this is reasonable for
  $K$, the width of $K^\star$ is $\Gamma_{K^\star} = 0.05\,$GeV,    
\item  $P$-wave tail suppression. In the VPP case the $K^\star$ decays
  to $K\pi$ in a $P$-wave, which is propagated to the $f_0\pi$
  final state, 
  $\left|\mathbb{M}_{a_1\to\pi f_0}^{(\text{vpp})}\right|^2 \sim 
  \left( \mathbb{A}_{a_1\to\pi f_0}^{(\text{vpp})} \right)^2
  \left|\vec{p}_{\pi}\right|^2$, ($\left|\vec{p}_{\pi}\right|$ is $a_1 \to f_0 \pi$ break up
  momentum). Therefore, the final $f_0$ and $\pi$ effectively are in
  a $P$-wave. This gives rise to an enhanced, unphysical tail in the
  signal intensity. 
\end{enumerate}

To take into account the finite width of intermediate particles 
we substitute propagators of stable particles by resonance propagators.
Technically this leads to substitutions $m_j^2 \to m_j^2 - i m_j \Gamma_j$.
Including such a term for the $K^\star$ propagator in
\eqnref{eq:Mvpp} results in a smoother behavior of the amplitude, as
shown in 
\figref{fig:ReIm} (right panel). The singularity at $\sqrt{s} = E_1 =
1.42\,$GeV is now limited and proportional to
$\log\Gamma_{K^\star}$. 

For a two-body decay with orbital angular momentum $L$, the
amplitude behaves like $p^L$ 
close to threshold due to centrifugal barrier. Far away from
threshold, this is no longer correct because of the finiteness of the
strong interaction. 
Accounting for the finiteness of interaction, however, is not
unique. For a direct decay of a resonance phenomenological form
factors are usually used, which 
come from a classical potential model. These could be Blatt-Weisskopf
barrier factors \cite{Blatt:1952} or exponential
factors for finite meson-size corrections \cite{Asner:1999kj}. Another
phenomenological approach is to introduce a simple left-hand singularity in
the amplitude as a vertex form factor \cite{Anisovich:1997pe}. We demonstrate here that the latter approach, where a pole is introduced in the amplitude to
account for the $K^\star$ $P$-wave decay,
gives a reasonable result. To do so, we include in \eqnref{eq:Mvpp} a factor   
\begin{equation}
F(k_1) = \frac{C}{M^2 - k_1^2}  
\end{equation}
under the integral, 
where $M$ is the position of the left-hand singularity and $C$ is a 
constant normalized to the $K^\star\to K \pi$ decay from mass shell, $C =
M^2 - m_{K^\star}^2$. 
Above the $K \pi$ threshold this correction behaves like a $D$-wave
Blatt-Weisskopf factor ($F_{\text{bw}}^{(D)}$). 
So $M$ corresponds to the size of $K^\star$
\begin{multline} \label{eq:BW}
F_{\text{bw}}^{(D)}(\vec{p})^{-1/2} \sim 1 + R^2 |\vec{p} \,|^2 \approx 
1+R^2 \frac{(k_1^2 - (m_\pi+m_K)^2)(k_1^2 - (m_\pi-m_K)^2)}{4k_1^2}
\approx \\
\approx 
1+R^2 \frac{(k_1^2 - (m_\pi+m_K)^2)}{4} = -\frac{R^2}{4} (M^2 - k_1^2), \quad 
M^2 = (m_\pi+m_K)^2 - \frac{4}{R^2}\enspace.
\end{multline}

With this form of correction, our standard approach of Feynman parameters and
Wick rotation can be used; the details of the calculation are again
moved to \appref{sec:adx.integral}. 
The final expression is  
\begin{multline} \label{eq:A.corr}
\mathbb{A}_{a_1\to f_0 \pi}^{(\text{vpp})} \to 
\frac{C}{16\pi^2} \int_0^1 \diff{y} \int_0^{1-y} \diff{z} \, \times \\
 \bigg(
\frac{(1-y-z)}{(\Delta_{yz}+m_1^2(1-y-z))(\Delta_{yz}+M^2(1-y-z))} +  \\
\frac{(z^2 p_1^2+yz(p_1\cdot p_2))(1-y-z)^2}{(\Delta_{yz}+m_1^2(1-y-z))\times(\Delta_{yz}+M^2 (1-y-z))\times\Delta_{yz}} - \\
-\frac{1}{4}\frac{2}{M^2}\left[
\frac{1}{m_1^2}\log\frac{\Delta_{yz}+m_1^2(1-y-z)}{\Delta_{yz}} - 
\frac{1}{m_1^2-M^2}\log\frac{\Delta_{yz}+m_1^2(1-y-z)}{\Delta_{yz}+M^2(1-y-z))}
\right]
\bigg)\enspace.
\end{multline}

\begin{figure}[tbp]
  \centering
  \includegraphics[width=0.49\textwidth]{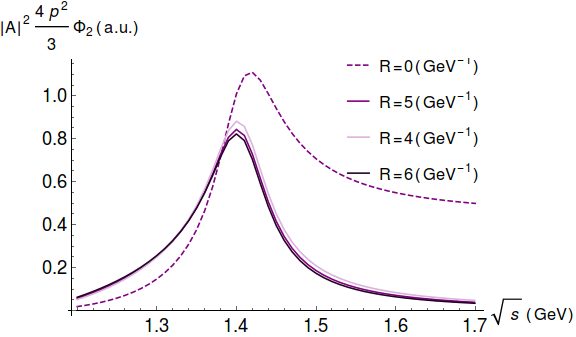} \
  \includegraphics[width=0.49\textwidth]{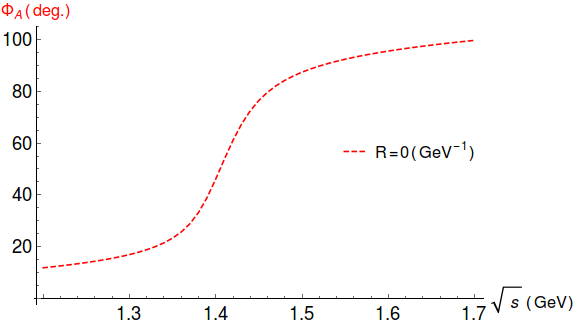}
  \caption{The intensity (left side) and the phase (right side) of the final amplitude
    $\mathbb{A}_{a_1\to\pi f_0}(s)$ including finite width of
    $K^\star$ and $P$-wave tail suppression 
    (see \eqnref{eq:vvp_case})  
    as a function of $\sqrt{s}$ for different values of the
    suppression parameter $R$.}  
  \label{fig:vpp.bw}
\end{figure}

The final result including the final width of $K^\star$ and the
suppression of the $P$-wave tail at higher energies is plotted in
\figref{fig:vpp.bw}. The left panel shows the intensity,
$\left|\mathbb{A}_{a_1\to\pi f_0}\right|^2 \left(4p^2/3\right) \, \Phi_2$, where $\Phi_2$ is the
two-body phase space, for  
different values of the size parameter 
$R = 0.8 - 1.2\,$fm, and also without suppression ($R=0$). The tail is
indeed suppressed as expected, almost independent of the exact value of $R$. The
phase of the signal for $R=0$ (no suppression) is shown in the right panel. 

Including the suppression factor in the integral of \eqnref{eq:Mvpp}
artificially shifts  
the phase to lower values with respect to the case with no
suppression. The phase motion, i.e.\ the relative difference as a
function of the energy is not affected. 
Therefore we show only the phase for $R=0$. 

\clearpage

\section{The reaction $\pi^- p\to a_1^-(1260) p \to f_0(980)\pi^- p$}
\label{sec:reaction}

\subsection{Cross section}
\label{sec:reaction.sigma}
\begin{figure}[ht]
\centering
\begin{tikzpicture}[node distance=1.5cm]
\coordinate[label=left:$p_p$] (a1);
\coordinate[	right=of a1,yshift=+2mm] (a2);
\coordinate[right=of a2,label=right:$p_4$,yshift=-2mm] (a3);
\coordinate[above=of a1,label=left:$p_\pi$,yshift=0mm] (b1);
\coordinate[right=of b1] (b2);
\coordinate[right=of b2,xshift=-3mm] (b3);
\coordinate[above right=of b3,yshift=-3mm] (b4);
\coordinate[above right=of b4,label=right:$p_1$,yshift=-4mm] (b6);
\coordinate[below right=of b4,label=right:$p_2$,yshift=+4mm] (b7);
\coordinate[below right=of b3,label=right:$p_3$,xshift= 0mm,yshift=5mm] (b5);

\draw[scalar]  (a1) -- node[label=above:$p$] {} (a2);
\draw[scalar]  (a2) -- node[label=above:$p$] {} (a3);
\draw[scalar]  (b2) -- node[label=right:$\mathbb{P}$] {} (a2);
\draw[pseudo]  (b1) -- node[label=above:$\pi^-$] {} (b2);
\draw[axial]   (b2) -- node[label=above:$a_1^-$] {} (b3);
\draw[pseudo]  (b3) -- node[label=below:$\pi^-$] {}  (b5);
\draw[vector]  (b3) -- node[label=above left:$R$] {}  (b4);
\draw[pseudo]  (b4) -- node[label=above:$\pi^-$] {}  (b6);
\draw[pseudo]  (b4) -- node[label=below:$\pi^+$] {}  (b7);
\end{tikzpicture}
\caption{Diagram for diffractive production of the $a_1(1260)$ by
  scattering of a high-energy $\pi^-$ off a proton target, which remains
  intact. The $\pi^-\pi^+\pi^-$ final state observed in the
  experiment is modeled by the decay of the $a_1^-$ into a charged pion
  and a neutral isobar $R$ ($\rho^0$ or $f_0(980)$), which subsequently decays
  into two charged pions. The isobar can be formed either by $\pi_1^-\pi_2^+$ 
  or by $\pi_2^+\pi_3^-$.}    
\label{fig:production}
\end{figure}
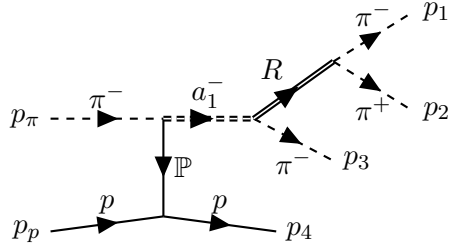

With a high-energy pion beam, as used in COMPASS and VES, the
$a_1^-(1260)$ is produced in a diffractive process proceeding via
$t$-channel Pomeron exchange between the beam $\pi^-$ and the target
proton, as shown in \figref{fig:production}.  

In order to estimate the intensity of the signal expected in the
$f_0(980)\pi^-$ channel we calculate its intensity and phase
difference compared to the dominant $a_1^-(1260)\to\rho^0\pi^-$ decay, assuming
that the signal in $f_0\pi^-$ is entirely due to the triangle singularity in the
decay $a_1^-(1260)\to f_0(980)\pi^-$. 
We denote the invariant mass squared of the $a_1$ by $s$, and the
isobar invariant mass squared by $s_{12}$ or $s_{23}$, respectively. 
Factorizing out the production cross section $\sigma_\mathrm{prod}(s)$
of the $a_1(1260)$, which 
is independent of the final state, the differential
cross section for the full process 
$\pi^- p\to a_1^-(1260) p\to R \,\pi^- p\to \pi^-\pi^+\pi^- p$
can be written as 
\begin{multline}  \label{eq:sigma_full}
\frac{\diff{\sigma}}{\diff{s}} 
= \frac{\sigma_{\text{prod}}(s)}{4\pi}
\left[
\int \, \frac{\diff{s_{12}}}{2\pi} \frac{2m_{a_1}\Gamma_{a_1^-\to
    R\pi^-}(s,s_{12})}{(m_{a_1}^2-s)^2+m_{a_1}^2\Gamma_{a_1}^2(s,s_{12})}  \,
\frac{2m_{R}\Gamma_{R\to\pi^+\pi^-}(s_{12})}{(m_{R}^2-s_{12})^2+m_{R}^2\Gamma_{R}^2(s_{12})}
+ \right. \\ 
\left. + \{12 \leftrightarrow 23\} + \int \diff{\Phi_{3\pi}} \,
  \text{Interf}(s_{12},s_{23}) 
\right]\enspace,
\end{multline}
where $\Gamma_{a\to bc}$ is the partial width for the two-body decay
$a\to bc$, $m_{a_1,R}$ and $\Gamma_{a_1,R}$ are the pole masses and 
mass-dependent full widths of $a_1$ and $R$, respectively, and
$\diff{\Phi_{3\pi}}$ is the $3\pi$ phase space. 
The first two terms constitute the contributions with the isobar
in the $\pi_1^- \pi_2^+$ and in the $\pi_2^+ \pi_3^-$
subsystem, respectively. The third term is the 
contribution of the interference between two processes. 
The latter is found to be very small for $a_1^- \to 
\rho^0 \pi^-$ (less then $2\%$ for $a_1$ decay from its mass shell) as well
as for $a_1^-\to f_0 \pi^-$, so we will disregard that contribution and
use the following equation for the cross section: 
\begin{equation}  \label{eq:sigma2}
\frac{\diff{\sigma}}{\diff{s}} \approx 
\frac{\sigma_{\text{prod}}(s)}{2\pi}
\int \, \frac{\diff{s_{12}}}{2\pi} \frac{2m_{a_1}\Gamma_{a_1^-\to R\pi^-}(s,s_{12})}{(m_{a_1}^2-s)^2+m_{a_1}^2\Gamma_{a_1}^2(s,s_{12})} 
\,
\frac{2m_{R}\Gamma_{R\to\pi^+\pi^-}(s_{12})}{(m_{R}^2-s_{12})^2+m_{R}^2\Gamma_{R}^2(s_{12})} \enspace.
\end{equation}

The partial widths of $a_1$ and isobar decays are calculated by
averaging the expressions for the square of the matrix elements of the corresponding
hadronic vertices 
over the initial
states and summing over the final ones and multiplying by the
corresponding phase space.   
For the decay $a_1^-(1260)\to\rho^0\pi^-$ (axial vector to vector and
pseudoscalar, AVP), we have 
\begin{equation}
\Gamma_{a_1^-\to\rho^0\pi^-}(s,m^2) = \frac{1}{2m_{a_1}} \, g_{a_1^- \rho^0
  \pi^-}^2\left[ 1+ \frac{|\vec{p}_{\rho}|^2}{3 m^2}\right] \times
\frac{1}{8\pi} \frac{2|\vec{p}_{\rho}|}{\sqrt{s}}\enspace, 
\end{equation}
while for $a_1^-(1260)\to f_0\pi^-$ (axial vector to pseudoscalar and
scalar, APS), we get  
\begin{equation}
\label{eq:Gamma}
\Gamma_{a_1^-\to f_0 \pi^-}(s,m^2) = \frac{1}{2m_{a_1}} \, g_{a_1^-
  f_0 \pi^-}^2(s,m^2) \frac{4|\vec{p}_{f_0}|^2}{3} \times
\frac{1}{8\pi} \frac{2|\vec{p}_{f_0}|}{\sqrt{s}}\enspace, 
\end{equation}
where $|\vec{p}_{\rho/f_0}(s,m^2)| =
\lambda^{1/2}(s,m^2,m_\pi^2)/(2\sqrt{s})$ is the break-up momentum for
the two-body 
decay of a particle with mass $\sqrt{s}$ to particles with masses $m=\sqrt{s_{12}}$
and $m_\pi$. The coupling of $a_1^-$ to $f_0\pi^-$ in \eqnref{eq:Gamma} is
given by 
\begin{equation}
g_{a_1^- f_0 \pi^-}^2(s,m^2) = \left|\mathbb{A}_{a_1\to f_0 \pi}^{(\text{vpp})}\right|^2 \,
\left(g_{a_1^- f_0 \pi^-}^{(K^\star \bar{K}+c.c.)} \right)^2\enspace, 
\end{equation}
where $\mathbb{A}_{a_1\to f_0 \pi}^{(\text{vpp})}$ is 
the triangle amplitude calculated in  
\eqnref{eq:A.corr} 
and $g_{a_1^-f_0 \pi^-}^{(K^\star\bar{K}+c.c.)}$  is an effective
coupling which 
includes the couplings of the individual vertices in the triangle
diagram, taking into account both isospin channels.  

The expressions for the isobar decays are 
\begin{equation}
\Gamma_{\rho^0\to\pi^+\pi^-}(m^2) = \frac{1}{2m_\rho} \, g_{\rho^0 \pi^+ \pi^-}^2
\frac{4|\vec{p}_{\pi}|^2}{3} \times \frac{1}{8\pi}
\frac{2|\vec{p}_{\pi}|}{m}\enspace,
\end{equation}
and 
\begin{equation}
\Gamma_{f_0\to\pi^+\pi^-}(m^2) = \frac{1}{2m_{f_0}}\,\frac{2}{3} g_{f_0\pi\pi}^2 \times
\frac{1}{8\pi} \frac{2|\vec{p}_{\pi}|}{m} \approx \frac{2}{3}\,\bar{g}_{f_0 \pi\pi}
|\vec{p}_{\pi}|\enspace, 
\end{equation}
where the dimensionless coupling $\bar{g}_{f_0\pi\pi} =
g_{f_0\pi\pi}^2/(8\pi m_{f_0}^2)$ has been introduced, and  
$\left|\vec{p}_{\pi}\right| = \lambda^{1/2}(m^2,m_\pi^2,m_\pi^2)/(2m)$ is
the break-up momentum of the isobar with mass $m$ to two pions. 

\subsection{Evaluation of the couplings}
\label{sec:reaction.coupling}
To evaluate the magnitude of $a_1\to f_0 \pi$ decay with respect to $a_1\to
\rho \pi$ $S$-wave we take into account  
two possible isospin configurations of intermediate states
($K^{\star 0} K^- K^+$) and ($K^{\star-} K^0 \bar{K}^0$) and evaluate
the corresponding couplings.
The table gives the couplings and Clebsch-Gordan coefficients for each vertex inside the loop of the two isospin configurations.
\begin{center}
\begin{tabular}{c | c | c }
Vertex & $K^{\star 0} K^- K^+$  & $K^{\star -} K^0
\bar{K}^0$ \\
\hline
\hline
$a_1^-$   & $g_{a_1 K^{\star} \bar{K}}$ & $g_{a_1 K^{\star} \bar{K}}$ \\
$\pi^-$ & $\sqrt{2/3}\,g_{K^{\star} K \pi}$ & 
$\sqrt{2/3}\,g_{K^{\star} K \pi}$\\
$f_0$   & $\sqrt{1/2}\,g_{f_0 K \bar{K}}$ & $\sqrt{1/2}\,g_{f_0 K \bar{K}}$\\
\end{tabular}
\end{center}
Since the isospin structure of both configurations is identical, the two diagrams add up. Disregarding
the mass difference between charged and neutral kaons, 
the contributions of both are the same, so the 
effective coupling of process can be written as  
\begin{equation}
g_{a_1^- f_0 \pi^-}^{(K^\star \bar{K}+c.c.)} = \frac{2}{\sqrt{3}} g_{a_1 K^{\star} K} \,g_{K^{\star}K\pi}\,
  g_{f_0 K \bar{K}} \enspace.
\end{equation}

We first consider $a_1$ decays. The resonance is rather wide, so the energy
dependence of the width should be taken into account. The best knowledge
about $a_1$ decay channels and branching ratio comes from hadronic
$\tau$ decay measurements \cite{Asner:1999kj, Briere:2003fr,
  Coan:2004ep}.  
For simplicity we consider only the main contribution to the energy dependence
of the width, which comes from $a_1\to \rho \pi$ $S$-wave.
\begin{equation}
  \Gamma_{a_1}(s) = \Gamma_{a_1}(m_{a_1}^2) \frac{|\vec{p}(s)|}{|\vec{p}(m_{a_1}^2)|} \frac{m_{a_1}}{\sqrt{s}}, \quad 
  |\vec{p}(s)| = \frac{\lambda^{1/2}(s,m_{\rho}^2,m_{\pi}^2)}{2\sqrt{s}}\enspace.
\end{equation}
We use the measured branching ratios, $\mathrm{Br}(a_1\to\rho\pi$,
$S-\text{wave})\approx 60\%$, $\mathrm{Br}(a_1\to K^\star \bar{K}+c.c.$,
$S-\text{wave})\approx 2.2\%$ to extract the ratio of couplings. 
The ratio $g_{a_1 \rho \pi}/g_{a_1 K^{\star}\bar{K}}$ is
calculated with the help of \eqnref{eq:sigma2}, where the energy
dependence of the production mechanism is disregarded. 
For $a_1\to\rho\pi$, a size correction form factor $\exp(-R^2 |\vec{p}_\rho\,|^2)$, where $\vec{p}_\rho$ the is break up momentum, is applied \cite{Asner:1999kj}. 
For $R>0.5\,$GeV$^{-1}$, the convergence of the integral over $\mathrm{d}s$ in \eqnref{eq:sigma2} is achieved for an upper limit of the integration $\leq 5\,$GeV, 
while for $R=0$ a higher limit is required.
Varying $R$ between $0$ and $5\,$GeV$^{-1}$, and including the uncertainty due to the slow convergence for $R=0$, the resulting ratio of the couplings is 
\begin{equation} \label{eq:rho.kstar}
\frac{g_{a_1 \rho \pi}^2}{g_{a_1 K^\star \bar{K}}^2} = 
\frac
  {2g_{a_1^- \rho^0 \pi^-}^2}
  {g_{a_1^- K^{\star 0} K^-}^2 + g_{a_1^- K^{\star -} K^0}^2} \approx 6-10\enspace.
\end{equation}
For the evaluation of relative strength of $f_0 \pi$ signal in \secref{sec:reaction.ratio}, we use $g_{a_1 \rho \pi}^2/g_{a_1 K^\star \bar{K}}^2 = 6$.

The width of $K^\star$ is measured precisely
and the branching to $K\pi$ $P$-wave is $100\%$ \cite{Agashe:2014kda}. 
The corresponding coupling can thus be extracted from 
\begin{equation}
\Gamma_{K^\star} = \frac{1}{2m_{K^\star}} \, g_{K^\star K\pi}^2 \frac{4|\vec{k}|^2}{3} \times \frac{1}{8\pi} \frac{2|\vec{k}|}{m_{K^\star}}, 
\quad
|\vec{k}| = \lambda^{1/2}(s,m_{K^\star}^2,m_\pi^2)/(2m_{K^\star}) 
\end{equation}
to be $g_{K^\star K \pi}^2 = 31.2$. 

The parameterization of the $f_0$ is not trivial, since both decay
channels ($f_0 \to 2\pi$, $f_0 \to 2K$) need to be taken into account. We make
use of the Flatt\'{e} parameterization \cite{Flatte:1976xu} of the $f_0$ propagator and the decay width in \eqnref{eq:sigma2},
\begin{equation}
\frac{ 2 m_{f_0}   \,\bar{g}_{\pi\pi} |\vec{p}_{\pi}|}
{\left| m_{f_0}^2-s_{12} - i m_{f_0} 
(\bar{g}_{\pi\pi} |\vec{p}_{\pi}|+\bar{g}_{K\bar{K}} |\vec{p}_{K}|) \right|^2 }, 
\quad 
\left| \vec{p}_{\pi/K}(s_{12}) \right| = \frac{1}{2}\sqrt{s_{12}-4m_{\pi/K}^2}\enspace.
\end{equation}
The measurements of the branching ratio $\text{Br}(f_0\to K\bar{K})/\left(\text{Br}(f_0\to\pi\pi)\right)$ and
the ratio of couplings extracted therefrom, 
$R_{K/\pi} = g_{f_0 K\bar{K}}^2/g_{f_0\pi\pi}^2 =
\bar{g}_{f_0K\bar{K}}/\bar{g}_{f_0\pi\pi} \approx 4$,
are rather consistent with each other \cite{Baru:2004xg,Ablikim:2004wn}.
But the absolute values of the couplings are not known very well. 
For our estimation of the branching we use $g_{f_0\pi\pi} = 2.3\,$GeV
\cite{GarciaMartin:2011jx}, so $\bar{g}_{f_0\pi\pi} = 0.21$,
$\bar{g}_{f_0 K\bar{K}} \approx 0.8$. 

\subsection{Evaluation of the branching ratio}
\label{sec:reaction.ratio}

\begin{figure}[tbp]
  \centering
  \includegraphics[width=0.49\textwidth]{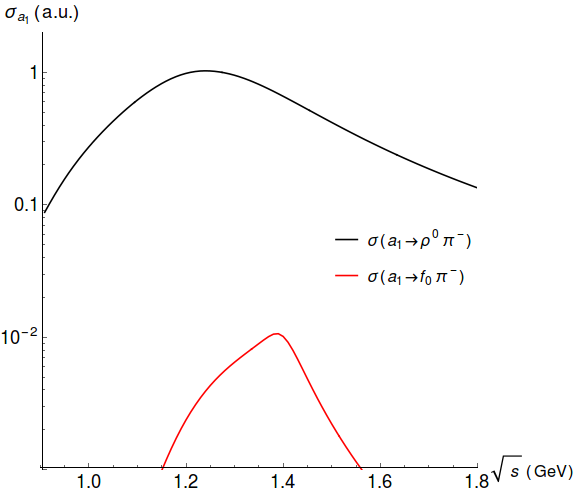}
  \includegraphics[width=0.49\textwidth]{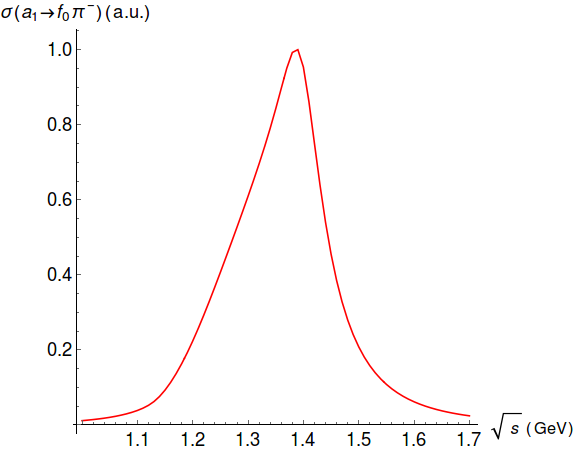}
  \caption{Cross sections for $a_1(1260)$ resonance as a function of
    invariant mass $\sqrt{s}$ in arbitrary units. (Left) comparison of dominant  
    $a_1^-\to\rho^0\pi^-$ $S$-wave decay and $a_1^-\to f_0 \pi^-$
    $P$-wave channel due to rescattering of kaons. (Right) Pseudo-resonant
    shape of $a_1^-\to f_0\pi^-$ in linear scale.} 
  \label{fig:invMass}
\end{figure}

The cross section calculated with the help of \eqnref{eq:sigma2} for
$a_1^-(1260) \to f_0 \pi^- \to \pi^-\pi^+\pi^-$ is compared to the one
for the dominant channel $a_1^-(1260) \to \rho^0 \pi^- \to
\pi^-\pi^+\pi^-$ in \figref{fig:invMass}. Here, 
the peak of $a_1 \to \rho \pi$ has been normalized to $1$, the $f_0 \pi$
channel is shown in relative scale. Under the assumptions detailed in
the previous subsection for the couplings, the peak-to-peak ratio is
$\approx 1:100$, in very good agreement with the experimental result.    


Let us discuss how reliable our estimation is and which factors
could affect the magnitude and the shape of $f_0 \pi$ peak. 
First of all, we assumed that the  
origin of both $K^\star \bar{K}$, which are  
rescattered to $f_0\pi$, and $\rho \pi$ is decay of the $a_1(1260)$ resonance.
In a hadron fixed-target experiment like COMPASS or VES, however,
there may be 
other processes which contribute to the same final state, e.g.\
non-resonant Deck-like 
processes \cite{Ascoli:1974hi}. 
We expect a rather large contribution of Deck-like background to
$\rho\pi$ $S$-wave signal \cite{Ascoli:1974hi} as well as to $K^\star
\bar{K}$ channel \cite{Berdnikov:1994kc}. 
Taking into account these processes, the resulting ratio of couplings 
could be different from \eqnref{eq:rho.kstar}. 

Secondly, $\rho \pi \to f_0 \pi$ rescattering plays an important role.  
Using the same method as in \secref{sec:triangle.vpp} one can show that
the triangle diagram $a_1 \to \rho \pi \to f_0 \pi$ gives a 
rather flat amplitude with a constant phase (magnitude $\approx 4\%$ of
peak intensity of $K^\star \bar{K}+c.c. \to f_0 \pi$). 
It interferes with the signal from $K^\star \bar{K}+c.c.$ and changes
its intensity and phase. Taking into account this contribution is in
principle possible, but 
requires the knowledge of the relative sign between $a_1\to\rho \pi$
and $a_1\to K^\star\bar{K}$, i.e.\ the relative sign of $s\bar{s}$ in 
$a_1$, which is unknown.   

The third uncertainty comes from the shape of the $f_0$ and the
corresponding coupling constants. We
found that the shape of our signal is stable for different values 
of $g_{f_0\pi\pi}$ and $R_{K/\pi}$.  
The relative intensity, however, is proportional to $g_{f_0\pi\pi}^2
R_{K/\pi}$, which could therefore easily change by a factor of two
depending on the input values.

\clearpage

\section{Conclusions}
\label{sec:conclusion}

Even after many years of intense studies, both experimentally and
theoretically, the excitation spectrum of hadrons is still not
understood. This is especially true in the region of charm and bottom
quarks, but also the light-quark sector sometimes bears
surprises.   

Present-day experiments are collecting extremely large event samples,
with allow them to perform analyses with very small statistical
uncertainties and permit them to find small signals which were not
observable before. Recently, 
the COMPASS experiment has reported the observation of a  
resonance-like signal with axial-vector quantum numbers
$J^{PC}=1^{++}$ in a completely unexpected 
mass region only about $0.2\,\GeV$ above the ground state
$a_1(1260)$, decaying to $f_0(980)\pi$. 

In this paper we show that a resonance-like signal with a maximum
intensity at $1.4\,\GeV$, compatible with the experimental result, can
be generated dynamically via a triangle 
singularity in the decay of the ground state $a_1(1260)$ to $K^\star
\bar{K}\text{ + c.c}$ and the subsequent rescattering of the $K$ from $K^\star$ decay to form 
$f_0(980)$. This process also generates a rather sharp phase motion, which
is not locked to the phase of the wide $a_1(1260)$. The singularity
appears in a kinematic region where the intermediate particles are collinear and on mass
shell. The structure of the amplitude is investigated in two ways:
first, the imaginary part is calculated using cutting rules;
second, the full amplitude is evaluated using Feynman rules in order
to obtain the imaginary and real parts. Both approaches are performed
for the hypothetical case of scalar intermediate particles and for the
realistic case of vector and pseudoscalar intermediate particles. It
is shown that both cases give very similar results. For the final
result of the triangle amplitude, we also include the finite width of
the $K^\star$ and a phenomenological factor to suppress the tail due
to the $P$-wave decay of the $K^\star$ at high energies. The inclusion
of this factor, however, is not unique, and also spoils the beauty of
the solution somewhat, because it shifts the phase of the
amplitude which is questionable and needs future investigation. The treatment of decays
with higher angular momenta inside loops in terms of analytical
solutions certainly needs more attention in the future.   

We then estimate the magnitude of the signal expected in the
$f_0(980)\pi$ channel due to the 
triangle singularity compared to the dominant decay of the $a_1(1260)$
to $\rho\pi$, also observed in the experiment. Our result gives a
relative peak intensity of $1\%$ for the $f_0\pi$ channel, with a
rather large uncertainty which is due partly to poorly known
couplings and partly to other rescattering processes like 
$a_1\to\rho\pi\to f_0\pi$.
The last process does not produce a singularity in the kinematically allowed region, 
but the corresponding amplitude interferes with the $K^\star \bar{K}$ amplitude and 
modifies its	 intensity and phase. 
In addition, we only consider the genuine $a_1(1260)$ resonance as source for
the triangle diagram, while it is known that in the reaction 
$\pi^- p \to 3\pi \, p$ there is a rather large contribution to the intensity 
in the $\rho\pi$ channel from non-resonant processes like the Deck effect, 
which may influence the relative branching ratio. 

The dynamical interpretation of the $a_1(1420)$
presented in this paper captures the main effect and probably 
accounts for a large fraction of the signal observed by COMPASS and
VES. 
As a next step, one may fit our amplitude to the data and compare
to the Breit-Wigner fit, and eventually extract better values for the
coupling constants. The data sample on the $a_1$ from 
$\tau$-decays should also be large enough to observe the 
$f_0 \pi$ peak if the data is fitted without phase locking of $f_0 \pi$
with $a_1$. 
In general, the large data samples available nowadays both
for light and heavy hadrons allow us to revisit effects which
were already discussed more than 30 years ago, but were almost
forgotten since then because data were too scarce to test them. These
may play an important role in our understanding of the hadron
spectrum. 


%
%


\section*{Acknowledgments}
\label{sec:acknow}
This work is supported by German Bundesministerium f\"ur Bildung und
Forschung as well as RNF grant $\#14-22-00281$. 
The authors would like to thank E.L.~Berger for valuable
discussions during Hadron2013, and especially Q.~Zhao and A.M.~Zaitsev for
independently pointing our attention to a triangle singularity as 
possible origin of the observation, and thus triggering the present
work.   
We also thank the COMPASS Collaboration and in particular
D.I.~Ryabchikov for numerous discussions on the COMPASS and VES
data.  

\appendix

\section{Parameterization of vertices and Feynman rules.}
\label{sec:adx.feynmann_rules}
First we mention the approach we use to parameterize the vertices for
interactions of particles. 
From symmetry considerations the Lorentz structure for vertices is the 
following (we use S = scalar, P = pseudoscalar, V = vector, A = axial vector): 

\begin{figure}[h]
\centering
\begin{tikzpicture}[node distance=0.8cm and 1.2cm]
\coordinate[label=left:{$\varepsilon_0,\,p_0$}] (a1);
\coordinate[right=of a1] (a2);
\coordinate[above right=of a2,label=right:$k_1$] (b1);
\coordinate[below right=of a2,label=right:$k_2$] (b2);
\coordinate[right=3 = of a2,label=right:{$=g_{VPP}\ \varepsilon_\mu (k_1-k_2)^\mu$}] (b3);

\draw[vector] (a1) -- node[label=above:$V$] {} (a2);
\draw[pseudo] (a2) -- node[label=above:$P$] {} (b1);
\draw[pseudo] (a2) -- node[label=below:$P$] {} (b2);
\end{tikzpicture}
\qquad
\begin{tikzpicture}[node distance=0.8cm and 1.2cm]
\coordinate[label=left:{$p_0$}] (a1);
\coordinate[right=of a1] (a2);
\coordinate[above right=of a2,label=right:{$k_1$}] (b1);
\coordinate[below right=of a2,label=right:$k_2$] (b2);
\coordinate[right=3 = of a2,label=right:{$=g_{SPP}$}] (b3);

\draw[scalar] (a1) -- node[label=above:$S$] {} (a2);
\draw[pseudo] (a2) -- node[label=above:$P$] {} (b1);
\draw[pseudo] (a2) -- node[label=below:$P$] {} (b2);
\end{tikzpicture}
\caption{Parameterization of VPP vertex ($P$-wave) and SPP vertex ($S$-wave)}
\end{figure}

\begin{figure}[h]
\centering
\begin{tikzpicture}[node distance=0.8cm and 1.2cm]
\coordinate[label=left:{$\varepsilon_0,\,p_0$}] (a1);
\coordinate[right=of a1] (a2);
\coordinate[above right=of a2,label=right:{$\varepsilon_1,\,k_1$}] (b1);
\coordinate[below right=of a2,label=right:$k_2$] (b2);
\coordinate[right=3 = of a2,label=right:{$=g_{AVP}\ \varepsilon_0^\mu\varepsilon_{1\mu}$}] (b3);

\draw[axial] (a1) -- node[label=above:$A$] {} (a2);
\draw[vector] (a2) -- node[label=above:$V$] {} (b1);
\draw[pseudo] (a2) -- node[label=below:$P$] {} (b2);
\end{tikzpicture}
\caption{Parameterization of AVP vertex ($S$-wave)}
\end{figure}

\begin{figure}[h]
\centering
\begin{tikzpicture}[node distance=1cm and 3cm]
\coordinate[] (a1);
\coordinate[right=of a1] (a2);
\coordinate[right=3 =of a2,label=right:{\Large $=\frac{1}{m_1^2-k_1^2-i \epsilon}$}] (a3);

\draw[pseudo] (a1) -- node[label=above:{$k_1,\,m_1$},label=below left:$P$] {} (a2);
\end{tikzpicture}
\qquad
\begin{tikzpicture}[node distance=1cm and 3cm]
\coordinate[] (a1);
\coordinate[right=of a1] (a2);
\coordinate[right=3 =of a2,label=right:{\Large $=\frac{g^{\mu\nu}-k_1^\mu k_1^\nu/k_1^2}{m_1^2-k_1^2- i \epsilon}$}] (a3);

\draw[vector] (a1) -- node[label=above:{$k_1,\,m_1$},label=below left:$V$] {} (a2);
\end{tikzpicture}
\caption{Propagators for pseudoscalar and vector particles}
\end{figure}


\section{Calculation of integrals}
\label{sec:adx.integral}

In this section we calculate three integrals:
\begin{equation} \label{eq:adx_int3}
V_3 = \int \frac{\mathrm{d} k_1^4}{(2\pi)^4\,i} 
\frac{1}
{\Delta_1 \Delta_2 \Delta_3},
\end{equation}
\begin{equation} \label{eq:adx_int4}
V_4 = \int \frac{\mathrm{d} k_1^4}{(2\pi)^4\,i} 
\frac{\varepsilon_{0\mu} 
\left( g^{\mu\nu} - \frac{k_1^\mu k_1^\nu}{k_1^2} \right) 
(p_{1} - k_3)_\nu}
{\Delta_1 \Delta_2 \Delta_3},
\end{equation}
\begin{equation} \label{eq:adx_int5}
V_5 = \int \frac{\mathrm{d} k_1^4}{(2\pi)^4\,i} 
\frac{\varepsilon_{0\mu} 
\left( g^{\mu\nu} - \frac{k_1^\mu k_1^\nu}{k_1^2} \right) 
(p_{1} - k_3)_\nu \times \frac{C}{M^2-k_1^2}}
{\Delta_1 \Delta_2 \Delta_3}.
\end{equation}
where $\Delta_1 = m_1^2 - k_1^2- i\epsilon$, $\Delta_2 = m_2^2 - (p_0-k_1)^2- i\epsilon$, $\Delta_3 = m_3^2 - (k_1-p_1)^2- i\epsilon$.

\subsubsection{First integral}
\label{sec:adx.integral.1}
For the calculation of $V_3$, Feynman parameters ($x,y,z$) are
introduced to rewrite the integral: 
\begin{equation}
V_3 = \int_0^1\int_0^1\int_0^1 \diff{x}\, \diff{y}\, \diff{z}\, 2! \,\delta(x+y+z-1) \,
\int \frac{\mathrm{d}^4 k}{(2\pi)^4\, i} \frac{1}{D^3}\quad,
\end{equation}
where $D = x(m_1^2 - k_1^2 - i\epsilon) + y(m_2^2 - (p_0-k_1)^2 - i\epsilon) + z(m_3^2 - (k_1-p_1)^2 - i\epsilon)$.
The quadratic form $D(k_1)$ can be reduced to diagonal form collecting
terms with $k_1$ and extracting the full square. The condition $x+y+z
= 1$ is used. 
\begin{equation} \label{eq:D3}
D = -(k_1 - y p_0 - z p_1)^2 + \Delta - i\epsilon,
\end{equation}
\begin{equation} \label{eq:delta_yz}
\Delta = xm_1^2 + \Delta_{yz}, \quad 
\Delta_{yz} = ym_2^2 + zm_3^2 - y(1-y-z)p_0^2 - z(1-z-y)p_1^2 - yz p_2^2.
\end{equation}
After shifting the variable of integration $k_1 \to l = k_1 - y p_0 -
z p_1$ we have  
\begin{equation} 
V_3 = \int_0^1\int_0^1\int_0^1 \mathrm{d} x\, \mathrm{d} y\, \mathrm{d} z\, 2! \, \delta(x+y+z-1) 
  \int\frac{\mathrm{d}^4 l}{(2\pi)^4\, i} \frac{1}{(-l^2+\Delta -
    i\epsilon)^3}\quad. 
\end{equation}

For the integration over $l_0$, notice that the denominator has poles when
$l_0^2 = \vec{l}^2 + \Delta^2 - i\epsilon$.  The positions of the
poles are functions of the external invariants $p_0^2$, $p_1^2$,
$p_2^2$ and the Feynman parameters. 
The basic idea, which we use is aimed to calculate the integral in the
region where $p_0^2 <0$, $p_1^2<0$, $p_2^2<0$, i.e. $\Delta>0$ for all
value of $x,y,z$. 
In that region we can rotate the contour of integration over $l_0$
anticlockwise (Wick rotation) and integrate along the imaginary axis. 
We make use of the transfer of the 
integration variable $l$ to Euler
space with integration variable $l_E$ ($l^2 = -l_E^2$), where the
integration is much simpler. 
One has:
\begin{equation}
V_3 = \int_0^1\int_0^1\int_0^1 \mathrm{d}x\, \mathrm{d}y\, \mathrm{d}z\, 2! \, \delta(x+y+z-1)
\int \frac{\mathrm{d}^4 l_E}{(2\pi)^4} \frac{1}{(l_E^2+\Delta-i\epsilon)^3}.
\end{equation}
The next step is integration over $l_E$. As a result we have:
\begin{equation} \label{eq:res_int3}
V_3 =
\frac{1}{16\pi^2} \int_0^1 \mathrm{d}y \int_0^{1-y} \mathrm{d}z
\frac{1}{\Delta_{yz}+m_1^2(1-y-z)-i\epsilon}.
\end{equation}
\Eqnref{eq:res_int3} is simple enough for numerical integration.

\subsubsection{Second integral}
\label{sec:adx.integral.2}
Let us consider integral $V_4$, \eqnref{eq:adx_int4}. 
The numerator can be simplified as 
\begin{equation} \label{eq:simpl_nom}
\varepsilon_{0\mu} 
\left( g^{\mu\nu} - \frac{k_1^\mu k_1^\nu}{k_1^2} \right) 
(p_{1} - k_3)_\nu = 
\varepsilon_{0\mu} \left( g^{\mu\nu} - 
\frac{k_1^\mu k_1^\nu}{k_1^2} \right) (2p_{1} - k_1)_\nu = 
2 (\varepsilon_0\cdot p_1) + 2\frac{(\varepsilon_0\cdot k_1) (k_1 \cdot p_1)}{-k_1^2}.
\end{equation}

One can notice that $k_1^2$ in the numerator has the same form as
$\Delta_0 = m_0^2 - k_1^2$ with mass $m_0^2 = 0$. The integral in
\eqnref{eq:adx_int4} is equal to  
\begin{equation} \label{eq:adx.sec.sum}
\frac{1}{2} V_4 = (\varepsilon_0\cdot p_1) \int \frac{\mathrm{d}^4 k_1 }{(2\pi)^4\, i} 
\frac{1}{\Delta_1 \Delta_2 \Delta_3 } + 
\int \frac{\mathrm{d}^4 k_1}{(2\pi)^4 \, i} 
\frac{(\varepsilon_0\cdot k_1) (k_1 \cdot p_1)}{\Delta_0 \Delta_1 \Delta_2 \Delta_3 }.
\end{equation}
The first integral in \eqnref{eq:adx.sec.sum} is equal to \eqnref{eq:res_int3}. 
For the second one we introduce four Feynman parameters:
\begin{multline}
\int \frac{\mathrm{d}^4 k_1}{(2\pi)^4 \, i} 
\frac{(\varepsilon_0\cdot k_1) (k_1 \cdot p_1)}{\Delta_0 \Delta_1 \Delta_2 \Delta_3 } = \\
\int_0^1 \mathrm{d} t \int_0^1 \mathrm{d} x \int_0^1 \mathrm{d} y \int_0^1 \mathrm{d} z \,
3! \, \delta(t+x+y+z-1)
\int \frac{\mathrm{d}^4 k_1}{(2\pi)^4 \, i} 
\frac{(\varepsilon_0\cdot k_1) (k_1 \cdot p_1)}{ D_4^4 },
\end{multline}
where for $D_4$ with condition $x_0+x_1+x_2+x_3 = 1$ we have the same
expression as \eqnref{eq:D3}.  
So the same shift of the integration variable $k_1$ is used, i.e. $k_1
\to l = k_1 - y p_0 - z p_1$. 

The expression in the numerator can be written as
\begin{equation}
(\varepsilon_0\cdot k_1) (k_1\cdot p_1) = l_\mu l_\nu \cdot \big[ \varepsilon_0^\mu p_1^\nu\big] + 
l_\mu \cdot \big[ (z p_1^2+y(p_1\cdot p_2)) \varepsilon_0^\mu + z (\varepsilon_0 \cdot p_1) p_1^\mu \big] + 
 z (\varepsilon_0 \cdot p_1) (z p_1^2+y(p_1\cdot p_2)
).
\end{equation}
After Wick rotation and the integration over angular variables
$\mathrm{d}\Omega_4$, the term proportional to $l_\mu$ gives zero and
$l_\mu l_\nu \to - g_{\mu\nu} \,l_E^2/4$.  
So one arrives at 
\begin{multline}
\int \frac{\mathrm{d}^4 k_1}{(2\pi)^4 \, i} 
\frac{(\varepsilon_0\cdot k_1) (k_1 \cdot p_1)}{\Delta_0 \Delta_1 \Delta_2 \Delta_3 } = 
\int_0^1 \mathrm{d} t \int_0^1 \mathrm{d} x \int_0^1 \mathrm{d} y \int_0^1 \mathrm{d} z \,
3! \, \delta(t+x+y+z-1) \times  \\
\times  (\varepsilon_0\cdot p_1) \left[ -\frac{1}{4} 
\int \frac{\mathrm{d}^4 l_E}{(2\pi)^4 } 
\frac{l_E^2}{ (l_E^2 + \Delta - i\epsilon)^4 } +
z(z p_1^2+y(p_1\cdot p_2))
\int \frac{\mathrm{d}^4 l_E}{(2\pi)^4 } 
\frac{1}{ (l_E^2 + \Delta - i\epsilon)^4 }
\right].
\end{multline}
All integrals converge. The integration over $\mathrm{d} t$ is removed
by a delta-function. 
\begin{multline} 
\frac{1}{2}\,V_4 = \frac{(\varepsilon_0\cdot p_1)}{16\pi^2} \left[ \int_0^1 \mathrm{d}y \int_0^{1-y} \mathrm{d} z  \frac{1}{\Delta_{yz} + m_1^2(1-y-z)-i\epsilon} + \right.\\
\left.
+ \int_0^1 \mathrm{d}y \int_0^{1-y} \mathrm{d}z 
\int_0^{1-y-z} \mathrm{d}x \left( \frac{yz(p_0\cdot p_1) + z^2 p_1^2}{(\Delta_{yz} + m_1^2 x-i\epsilon)^2} - \frac{1/4}{ \Delta_{yz} + m_1^2 x-i\epsilon }\right)
\right],
\end{multline}
$\Delta_{yz}$ is given by \eqnref{eq:delta_yz}.

\subsubsection{Third integral}
The calculation of $V_5$, \eqnref{eq:adx_int5} proceeds similarly to $V_4$. 
The difference is that we have four poles instead of three in the
denominator of \eqnref{eq:adx_int5}. 
First, one can simplify the numerator as in \eqnref{eq:simpl_nom}:
\begin{equation} 
\frac{1}{2 C} V_5 = (\varepsilon_0\cdot p_1) \int \frac{\mathrm{d}^4 k_1 }{(2\pi)^4\, i} 
\frac{1}{\Delta_1 \Delta_2 \Delta_3 \Delta_4} + 
\int \frac{\mathrm{d}^4 k_1}{(2\pi)^4 \, i} 
\frac{(\varepsilon_0\cdot k_1) (k_1 \cdot p_1)}{\Delta_0 \Delta_1 \Delta_2 \Delta_3 \Delta_4},
\end{equation}
where $\Delta_4 = M^2 - k_1^2$.

We introduce four and five Feynman parameters for the integrals, respectively. 
The expression for the denominators are $D_5^4$ and $D_5^{\prime 5}$:
\begin{equation}
D_5 = x (m_1^2 - k_1^2) + y (m_2^2 - (p_0-k_1)^2) + z (m_3^2 - (k_1-p_1)^2) + u (M^2 - k_1^2),
\end{equation}
\begin{equation}
D_5' = t (- k_1^2) + x (m_1^2 - k_1^2) + y (m_2^2 - (p_0-k_1)^2) + z (m_3^2 - (k_1-p_1)^2) + u (M^2 - k_1^2).
\end{equation}
Then we perform the same calculation as in
\secref{sec:adx.integral.2}, and the result is  
\begin{multline} 
\frac{1}{2C}\,V_5 = \frac{(\varepsilon_0\cdot p_1)}{16\pi^2} 
\int_0^1 \mathrm{d}y \int_0^{1-y} \mathrm{d} z \int_0^{1-y-z} \mathrm{d} x 
\left[ \frac{1}{(\Delta_{yz} + m_1^2 x + M^2(1-x-y-z)-i\epsilon)^2} + \right.\\
\left.
+ \int_0^{1-x-y-z} \mathrm{d} u
\left( \frac{yz(p_0\cdot p_1) + z^2 p_1^2}{(\Delta_{yz} + m_1^2 x + M^2 u -i\epsilon)^3} - \frac{1/4}{ (\Delta_{yz} + m_1^2 x + M^2 u -i\epsilon)^2 }\right)
\right],
\end{multline}
where $\Delta_{yz}$ is given by \eqnref{eq:delta_yz}.
To make the expression simpler and convenient for a numerical
evaluation, we carry out the integration over $\mathrm{d} x$ and
$\mathrm{d} y$ explicitly:  
\begin{multline} \label{eq:A.corr}
\frac{1}{2C}\,V_5 = 
\frac{1}{16\pi^2} \int_0^1 dy \int_0^{1-y}dz \, \times \\
 \bigg(
\frac{(1-y-z)}{(\Delta_{yz}+m_1^2(1-y-z)-i\epsilon)(\Delta_{yz}+M^2(1-y-z)-i\epsilon)} +  \\
\frac{(z^2 p_1^2+yz(p_1\cdot p_2))(1-y-z)^2}{(\Delta_{yz}+m_1^2(1-y-z)-i\epsilon)(\Delta_{yz}+M^2 (1-y-z)-i\epsilon)(\Delta_{yz}-i\epsilon)} - \\
-\frac{1}{4}\frac{2}{M^2}\left[
\frac{1}{m_1^2}\log\frac{\Delta_{yz}+m_1^2(1-y-z)-i\epsilon}{\Delta_{yz}-i\epsilon} - 
\frac{1}{m_1^2-M^2}\log\frac{\Delta_{yz}+m_1^2(1-y-z)-i\epsilon}{\Delta_{yz}+M^2(1-y-z)-i\epsilon)}
\right]
\bigg).
\end{multline}
\bibliographystyle{h-physrev}
\bibliography{hadron,compass}

\end{document}